\documentclass[twocolumn,twoside,slac_two]{revtex4}
\usepackage{graphicx}
\usepackage{fancyhdr}
\def\cnone{\widetilde \chi_1^0}
\def\mcnone{m_{\cnone}}

\begin{document}

\title{Light Higgses and Dark Matter at Bottom and Charm Factories}

%

\author{Bob McElrath}
\affiliation{University of California-Davis}

\begin{abstract}
Neither Dark Matter nor scalar particles in the Higgs sector are ruled
out at energies accessible to bottom and charm factories.  In Dark
Matter searches, the error on the mass of Dark Matter is $\sim 4$ GeV in
the best LHC studies.  For light Dark Matter this could represent a
100\% (or more) error.  In Higgs searches, the presence of a light singlet
Higgs can make the LHC Higgs search difficult, if not impossible.  If
Dark Matter or a Higgs scalar is light, it will {\it require} a
low-energy machine to precisely determine the couplings.  We review the
models, modes of discovery and rate expectations for these new particle
searches at bottom and charm factories.  We also discuss the options for
new runs at bottom and charm factories relevant for these searches.
\end{abstract}

\maketitle

\thispagestyle{fancy}


\section{Introduction}

The two major new particles expected at colliders are Dark Matter and
the Higgs boson.  While some models are now ruled out at energies
accessible to bottom and charm factories, it is by no means proven that
these {\it cannot} be light.  In fact there exist many attractive models
containing light Higgses, for instance supersymmetric models which solve
the $\mu$ problem via an extended Higgs sector.\cite{nmssmdm,Han:2004yd,Barger:2006dh}
Furthermore the problem
of light Dark Matter and light Higgses are related, as light Dark Matter
particle $\chi$, in its simplest incarnation, requires a new light
particle $U$ with $m_U \simeq 2 m_\chi$ to serve as an $s$-channel
annihilation mediator.  A promising possibility for $U$ is that it is a
pseudo-scalar higgs, which can be naturally light due to new symmetries which
can protect its small mass.\cite{nmssmdm}
In order to ensure discovery, we should look
everywhere that is practical for solutions to these problems, and $b$- and
$c$-factories can perform an important set of new-particle searches.

Apart from the Dark Matter question, in the MSSM it was rigorously shown
that an extremely light neutralino is experimentally unconstrained if
one drops the assumption of gaugino unification, and the requirement
that the neutralino relic density be equal to the Dark Matter relic
density.\cite{Dreiner:2007fw} For instance, the Dark Matter problem
could be solved in another manner, such as with the QCD axion, rendering
the neutralino an insignificant contributor to the relic density of the
universe.

\section{Dark Matter}
In Dark Matter searches, the error on the mass of Dark Matter is $\sim
4$ GeV at the LHC in the best studies using optimistic models with large
cross sections.\footnote{These studies all use a large value $\sim 100$
GeV for the Dark Matter mass.  As this mass is brought closer to zero,
the resolution on it at the LHC worsens.} Ultimately this is due to the
resolution of the hadronic calorimeter, since to determine the mass
scale, a missing energy event at the LHC can either use the missing
transverse momentum, which is a hadronic observable\cite{Cheng:2007xv}
or if purely leptonic observables exist, one can use the changes in
slopes and shapes as a function of overall mass scale, which are only
weakly correlated.\cite{Gjelsten:2004ki}

The only truly fundamental limit on the mass of Dark Matter comes from
the Cosmic Microwave Background, which tells us the fraction of the
universe that was non-relativistic at the time that photons decoupled, a
measurement which includes Dark Matter.  Dark Matter must have been
non-relativistic at a temperature of about $0.3$ eV, therefore the
smallest possible mass consistent with the Standard Cosmological Model
is about 0.3 eV.

This means that bottom and charm factories are capable of exploring 10
orders of magnitude in the Dark Matter mass.  The LHC can expand to the
range 5 GeV -- 1 TeV, but has no precision below approximately 4 GeV.

We feel that the most compelling motivation for Dark Matter searches at
bottom and charm factories is the demonstrable wisdom of a model
independent approach.  Indeed, the reason $M < 45$ GeV was ignored for
so long is due to heavy reliance on models.  In particular the Minimal
Supersymmetric Standard Model cannot support Dark Matter this light
because it would require another charged or colored particle to be
lighter than other limits.  The secondary particle is necessary to get
the annihilation cross section large enough.  Nearly all models which
cannot support light Dark Matter cannot do so because of limits on {\it
particles other than the Dark Matter candidate itself}.  Trivial
extensions of these models can generically support light Dark Matter by
adding a mediator which is mostly singlet under the Standard Model.
Several models demonstrate this explicitly.\cite{nmssmdm,Boehm:2003hm}

\begin{figure*}[t]
    \includegraphics[scale=0.62]{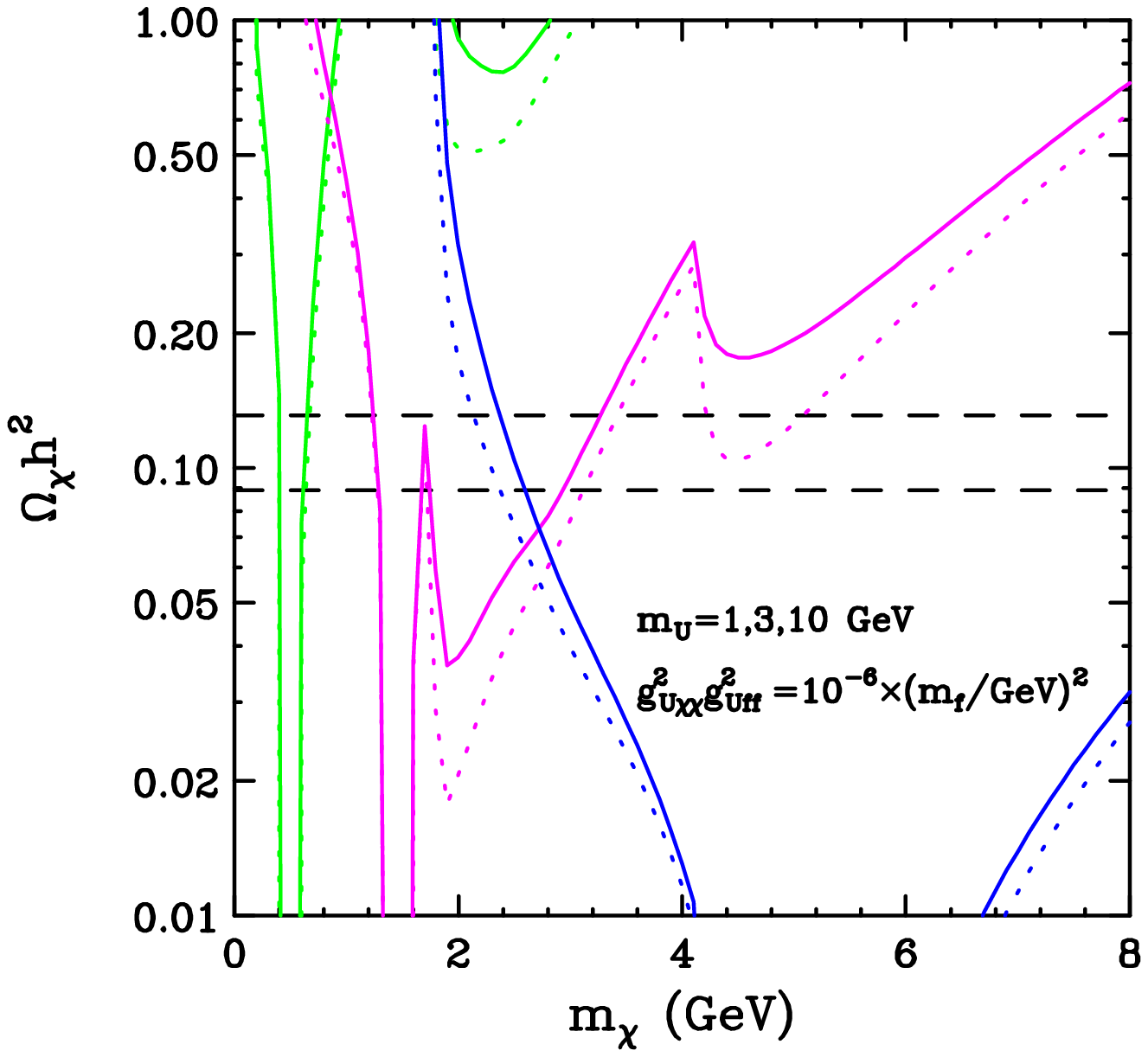}
    \includegraphics[scale=0.62]{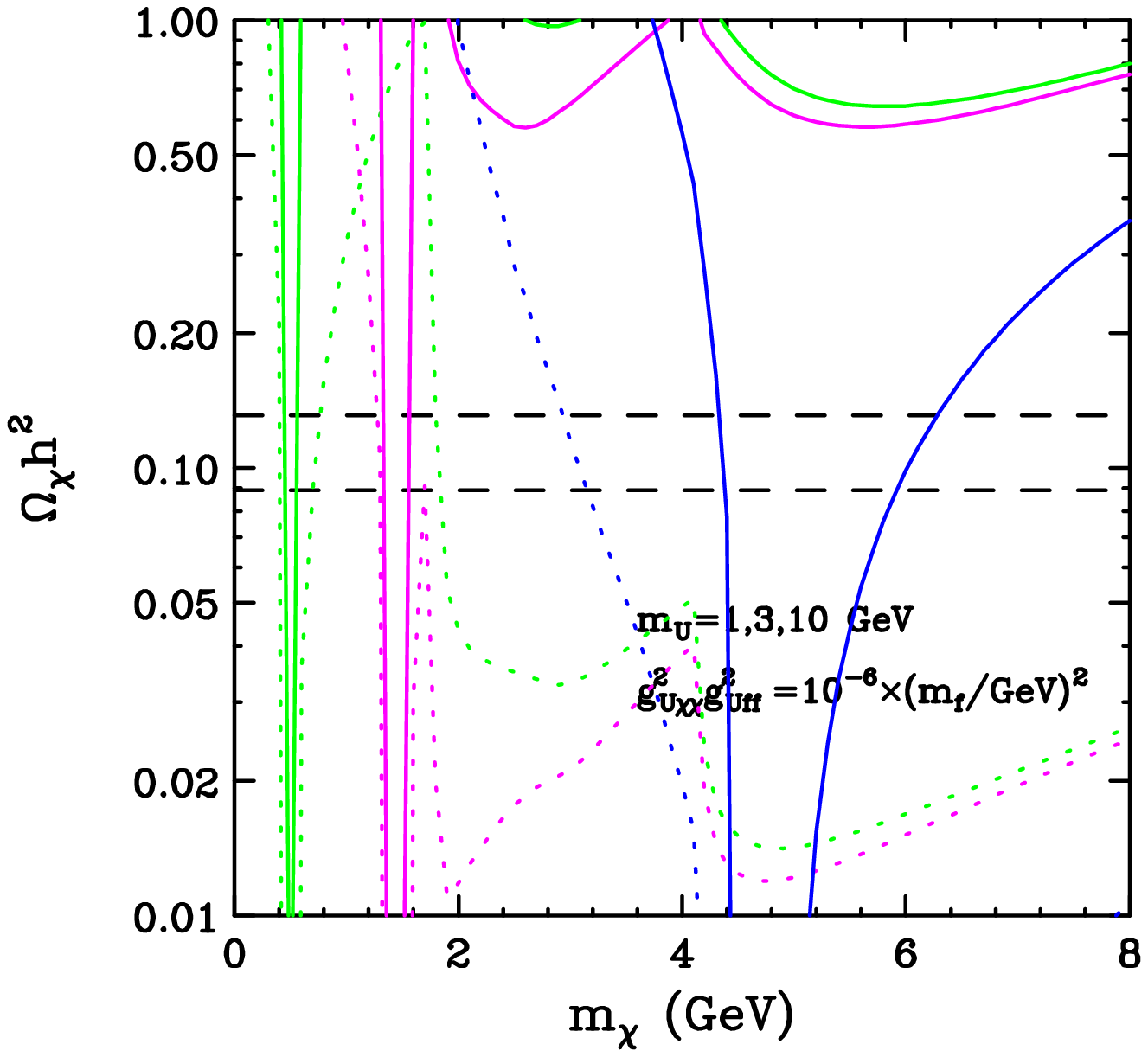}
    \centering
    \caption{The relic density $\Omega_\chi h^2$ vs the Dark Matter mass
    $M_\chi$ for masses of the mediator $M_U=1,3,10$ GeV.  In the left
    (right) panel, $\chi$ is a scalar (fermion).  In both panels, $U$ is a
    scalar (solid) or pseudo-scalar (dotted).  These curves move
    vertically as the free-parameter couplings $g_{Uff}$ and
    $g_{U\chi\chi}$ are changed.  We thank Dan Hooper for his
    contribution in creating these figures.}
    \label{omegah2mx}
\end{figure*}

The most minimal model possible for Dark Matter is to add only the dark
matter candidate $\chi$ itself.\cite{Cirelli:2005uq}  However these
models generate very heavy Dark Matter candidates, outside the reach of
$b$- and $c$- factories.  

The second most minimal models adds the mediator $U$ as well, which is
flavor neutral and couples both to the Standard Model and Dark Matter.
these are the models testable at $b$- and $c$- factories.

$U$ can be a new gauge boson as proposed in
Refs.\cite{Boehm:2003bt,Boehm:2003hm}, or a scalar as proposed in
\cite{nmssmdm}.  If $U$ is a vector, it is necessarily anomalous, so
building a consistent model requires even more matter than the $U$ and
$\chi$.  We are not aware of any such model in the literature.  This is
not because it is impossible, but rather because the resulting models
are ugly, requiring several symmetry breaking scales and associated
Higgses, as well as extra matter to cancel anomalies.  If $U$ is a
scalar, it can only couple to Standard Model fermions by mixing with the
Higgs bosons due to gauge invariance, making its couplings proportional
to mass.  In the author's opinion, a scalar or pseudo-scalar is a more
natural candidate for $U$.  Though as we will see in the next section, a
vector $U$ may be easier to discover.

Treating the relic density as a constraint, acceptable models are
achieved for (at least) two values of the Dark Matter mass as a function
of the mediator's mass, $M_\chi = M_U/2 \pm \epsilon$, as can be seen in
Fig.\ref{omegah2mx} This is because the process controlling the
annihilation of Dark Matter in the early universe is an $s$-channel
annihilation diagram, rather than t-channel diagrams.  In order for a
t-channel diagram to dominate the annihilation, the new particle in the
t-channel must be charged or colored.\footnote{For the masses we
consider, $M_\chi < 5$ GeV, annihilation to neutral Higgses or Z-bosons
is kinematically disallowed.}  This occurs in the MSSM (e.g. ``stau
co-annihilation'') and generates one of the most promising regions of parameter
space.

There is experimental evidence that Dark Matter may be light from the
INTEGRAL satellite\cite{Jean:2003ci}, which has detected an anomalously
large population of positrons in the galactic center, as suggested in
Ref.\cite{Boehm:2003bt}.  If this is from Dark Matter annihilation, it
requires $M_\chi \lesssim 3$ MeV.\cite{Beacom:2005qv}

Another source of evidence is from the DAMA annual modulation signal.
As shown in Ref.\cite{damadm}, this is consistent with light Dark Matter
due to the lower threshold of Sodium, as compared to heavier elements
such as gallium (CDMS) and xenon (XENON).

There are two major modes of discovery for light Dark Matter: invisible
meson decay\cite{McElrath:2005bp,Fayet:2007ua} and radiative
decay\cite{nmssmdm,Bird:2006jd}.  These are described respectively in
the following subsections.

\begin{figure*}[t]
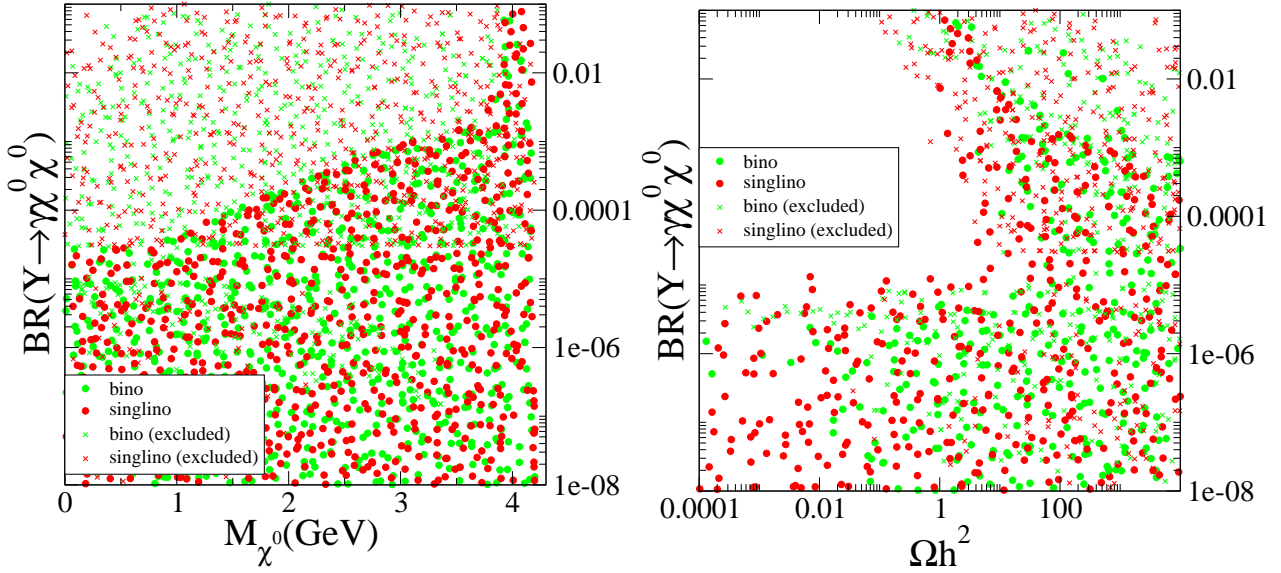

    \includegraphics[scale=0.33]{mn_brups.eps}
    \includegraphics[scale=0.33]{omegah2_brups.eps}
    \centering
    \caption{The branching ratio for $\Upsilon(1S)\to \gamma\cnone\cnone$  
    via 3-body decay (i.e. either $m_{A_1}<2\mcnone$ or
    $m_{A_1}>m_\Upsilon$) 
    is plotted vs.
    the LSP mass (left) and relic density $\Omega h^2$ (right). All points shown are
    consistent with all LEP constraints.  Points marked by an x are excluded by
    one of: $\Upsilon \rightarrow \gamma \cnone \cnone$ (3-body decay)
    (that which is plotted); $\Upsilon \rightarrow
    \gamma A_1$ (2-body decay)
    with  $A_1 \rightarrow \cnone \cnone$
    (2-body decay);  or
    $\Upsilon \rightarrow \gamma A_1$ (2-body decay) where the $A_1$ decays visibly.  }
\label{mn_brups}
\end{figure*}

\subsection{Invisible Quarkonium Decay}
In invisible meson decay, one can make a {\it na\"ive} calculation of
the branching ratio for a meson.\footnote{Calculation here is expanded
and corrected
relative to Ref.\cite{McElrath:2005bp}, however the uncertainties and
approximations made introduce much larger errors than the difference
between the two calculations.  This is an order-of-magnitude estimate
{\it only} and little significance can be attached to achieving or exceeding
these predictions.}  One can get an order of magnitude
estimate for the annihilation cross section using 
\begin{equation}
\Omega_X h^2 \simeq \frac{0.1 {\rm pb} \cdot c}
{\langle \sigma v \rangle } .
\end{equation}
Where $\Omega_X=\rho_X/\rho_c$ is the relic density for species $X$ relative to the
critical density $\rho_c$, $h$ is the Hubble constant, and
$\langle \sigma v \rangle$ is the thermally averaged annihilation cross
section of the DM into Standard Model particles.  
Using the central value of the
WMAP~\cite{wmap} result for $\Omega_X h^2 = 0.113$, we can invert this
equation and solve for the required annihilation cross section for light
relics
\begin{equation}
\langle \sigma v \rangle = 0.88\,{\rm pb}\cdot c.
\end{equation}
The velocity $v$ appearing here is the M\o ller velocity, which we approximate
by the relative velocity in the center-of-mass frame, $v_{\rm rel}=|v_1-v_2|$, using $\langle v_{\rm rel}^2 \rangle = 6/x_{FO}$.
The approximate temperature at freeze-out is $T=m_\chi/x_{FO}$ where
$m_\chi$ is
the mass of the DM and $x_{FO}$ is an expansion parameter
evaluated at the freeze-out temperature that is 
$x_{FO} \sim 20-25$ depending on the model.  By approximating that
$\langle \sigma v \rangle = \sigma \sqrt{\langle v_{\rm rel}^2\rangle}$ we can remove the
kinematic velocity factor, assuming that the per-particle energy is given by the
average energy of the gas ${3\over 2} k T$.

We can expand $\langle
\sigma v \rangle$ in the velocity at freeze-out to separate $s$-wave and
$p$-wave components,
$\langle \sigma v \rangle = a + b v^2$.  Since the
Dark Matter annihilates through the U-boson and not the meson we're
interested in, the freeze-out may in general occur at a different energy
than the invisibly-decaying meson mass.  Therefore we also remove the
extra $v^2$ term and solve for $b$ in the $p$-wave case.  These
manipulations remove the kinematic factors of the initial state, giving
us a cross section that essentially assumes the Dark Matter is massless
with respect to our invisibly-decaying meson; that the meson mass is
much larger than the center-of-mass energy at freeze-out.  

For these assumptions with $x_{FO}=25$ at freeze-out we have: 
\begin{eqnarray}
    \label{annxsect}
    \sigma(\chi \chi \rightarrow SM) = a/v_{\rm rel}&\simeq 1.8\,{\rm pb},&\quad {\rm (s-wave)} \\
    \nonumber
    \sigma(\chi \chi \rightarrow SM) = b/v_{\rm rel}&\simeq 7.5\,{\rm pb}. &\quad {\rm (p-wave)} 
\end{eqnarray}

The invisible branching ratio of a hadron can then be estimated by
assuming that the time-reversed reaction is the same, $\sigma(f \bar{f}
\rightarrow \chi \chi) \simeq \sigma(\chi \chi \rightarrow f \bar{f})$.
Since the meson decays by the meson mixing with the $U$ boson, $p$-wave
suppression factors are not reintroduced for the reverse reaction.
We assume that the DM mediator is not flavor changing and that
annihilation occurs in the $s$ channel.  Therefore, the best-motivated hadrons to have an invisible
width are same-flavor quark-antiquark bound states (quarkonia) with
narrow widths.

The invisible width of a hadron composed dominantly of $q
\bar{q}$ is given approximately by:
\begin{equation}
    \Gamma(H \rightarrow \chi \chi) = f_H^2 M_H \sigma(q \bar{q}
    \rightarrow \chi \chi) 
\end{equation}
where $f_H$ is the hadronic form factor (wave function at the origin)
for the state $H$, and $M_H$ is the hadron's mass.  Here we ignore
final state kinematic and spin factors.

We can predict an approximate expectation for the branching ratios for
narrow states.  Some of the most promising are shown in Table \ref{brs}:
\begin{table}
    \begin{tabular}{l| c |c}
    mode & $s$-wave & $p$-wave \\
    \hline
    BR($\Upsilon(1S) \rightarrow \chi \chi$)& $\ 4.2 \times 10^{-4}\ $& $\ 1.8 \times 10^{-3}\ $\\
    BR($\Upsilon(1S) \rightarrow \nu \bar{\nu}$)& $\ 9.9 \times 10^{-6}\ $&\\
    BR($J/\Psi \rightarrow \chi \chi$)      & $\ 2.5 \times 10^{-5}\ $& $\ 1.0 \times 10^{-4}\ $\\
    BR($J/\Psi \rightarrow \nu \bar{\nu}$)  & $\ 2.7 \times 10^{-8}\ $&\\
    BR($\eta \rightarrow \chi \chi$)        & $\ 3.4 \times 10^{-5}\ $& $\ 1.4 \times 10^{-4}\ $\\
    BR($\eta^\prime \rightarrow \chi \chi$) & $\ 3.7 \times 10^{-7}\ $& $\ 1.5 \times 10^{-6}\ $\\
    BR($\eta_c \rightarrow \chi \chi$)      & $\ 1.3 \times 10^{-7}\ $& $\ 5.3 \times 10^{-7}\ $\\
    BR($\chi_{c0}(1P) \rightarrow \chi \chi$)& $\ 2.7 \times 10^{-8}\ $&$\ 1.2 \times 10^{-7}\ $\\
    BR($\phi \rightarrow \chi \chi$)        & $\ 1.9 \times 10^{-8}\ $& $\ 7.8 \times 10^{-8}\ $\\
    BR($\omega \rightarrow \chi \chi$)      & $\ 7.2 \times 10^{-8}\ $& $\ 3.0 \times 10^{-8}\ $
    \end{tabular}
    \caption{Estimated branching ratios for the narrowest mesons.  The
    two columns correspond to the assumption that the Dark Matter
    annihilation in the early universe occurs in either the $s$-wave or
    $p$-wave.  Neutrino branching ratios are from Ref.\cite{Chang:1997tq}.  All
    mesons have a branching ratio (even if tiny) to neutrinos.}
    \label{brs}
\end{table}
Branching ratios for scalars and pseudo-scalars tend to be smaller since
those states are wider.  

We emphasize again that this is only an order-of-magnitude calculation.
A more precise calculation requires inclusion of kinematic and spin
factors, as well as consideration of which fermions the mediator $U$
couples to.  Furthermore, the freeze-out of light Dark Matter occurs in
the middle of the QCD phase transition, and is much more sensitive to
uncertainties due to QCD than heavier Dark Matter.  This kind of dark
matter is also annihilating through a narrow pole, which must be treated
carefully.\cite{Griest:1990kh}  Narrow poles arise due to the $U$ boson
itself, as well as numerous QCD resonances.

\begin{figure}[t]
    \label{brgammachichi}
    \includegraphics[angle=90,scale=0.4]{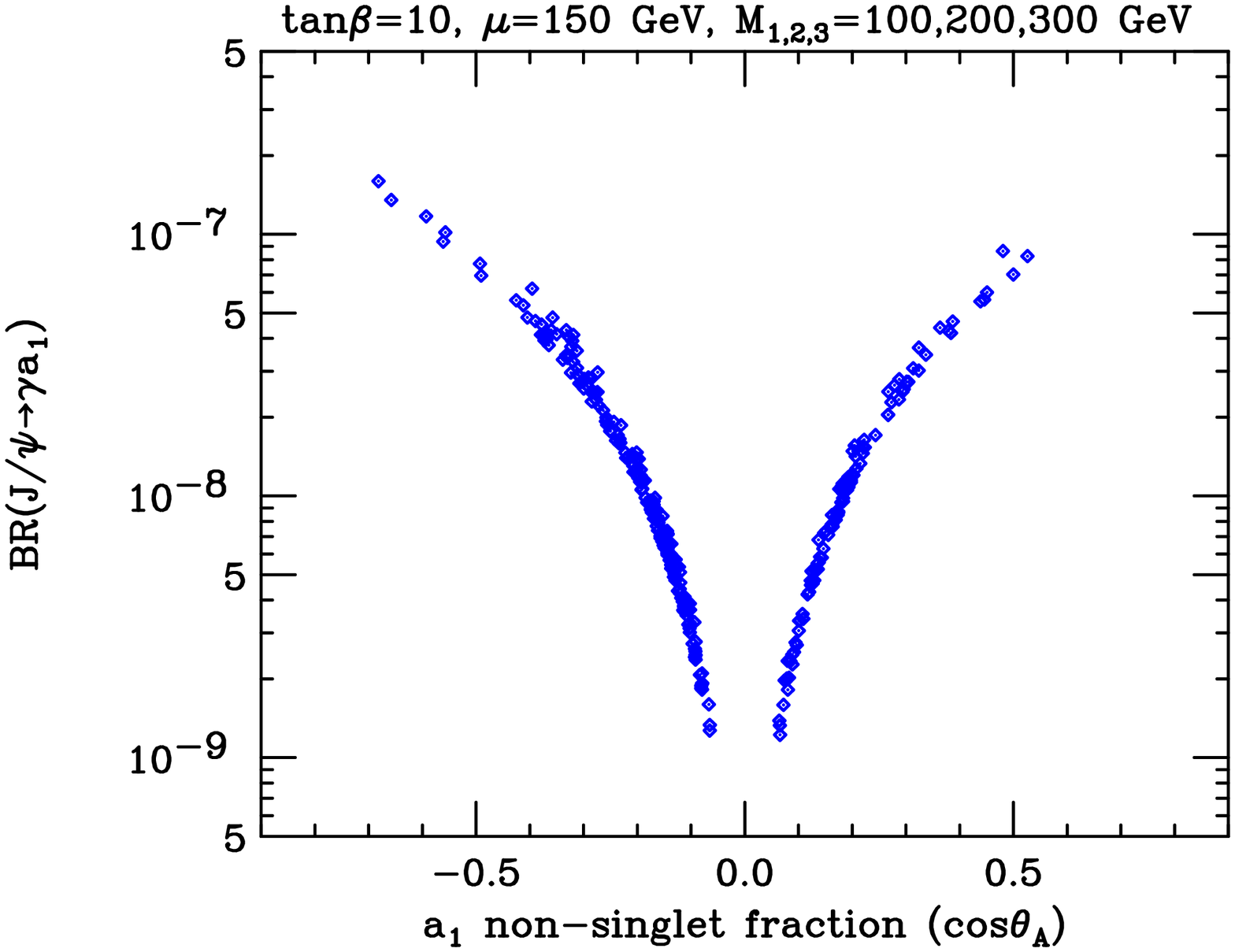}
    \includegraphics[angle=90,scale=0.4]{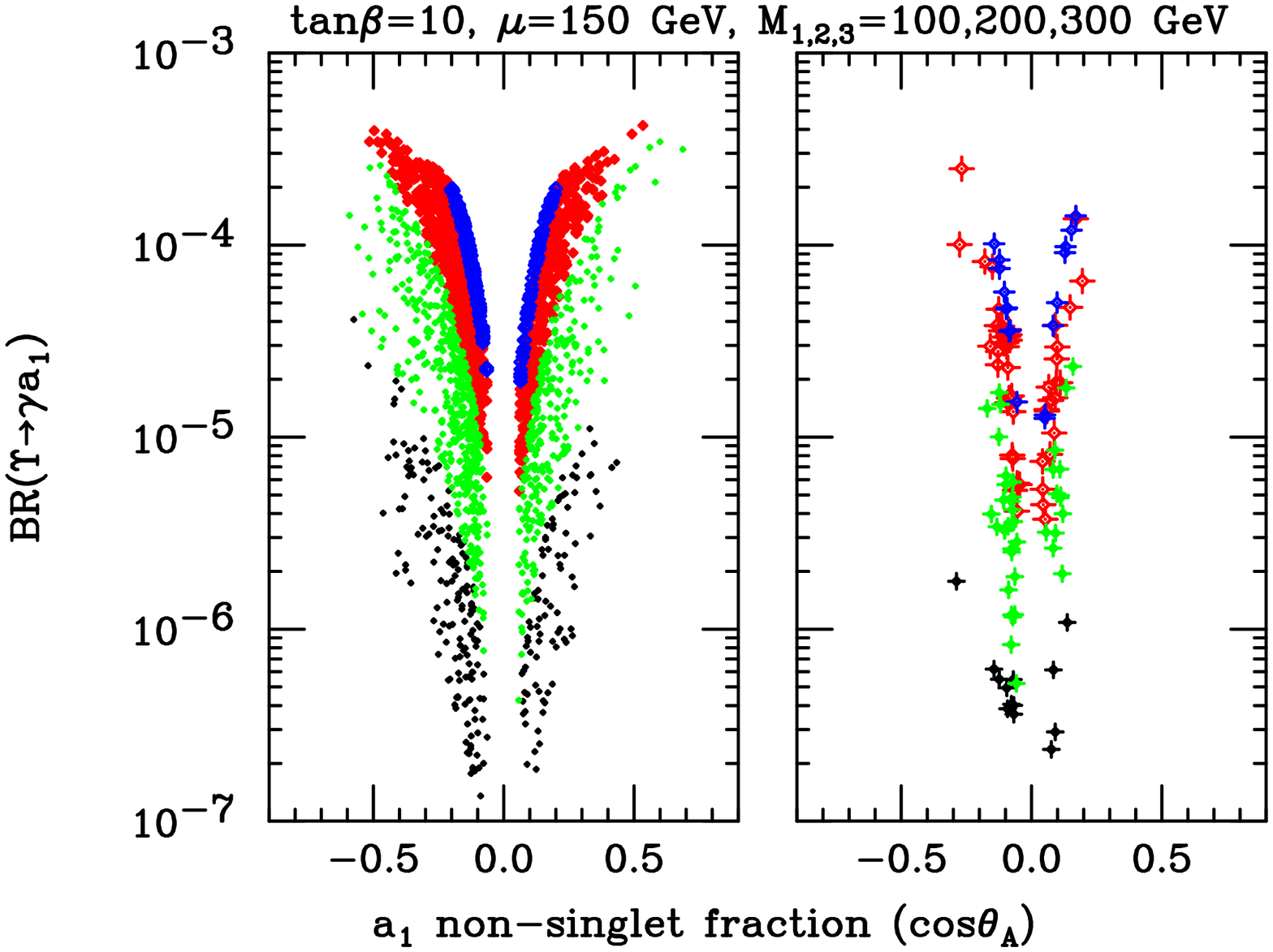}
    \centering
    \caption{Branching ratio of the $J/\Psi$ (top) and $\Upsilon$
    (bottom) into a photon and lightest pseudo-scalar Higgs $a_1$ in the
    NMSSM\cite{Dermisek:2006py}.  The $a_1$ may then decay into Dark
    Matter (neutralinos) or visible Standard Model particles.  The
    quantity $\cos \theta_A$ parameterizes how singlet-like the $a_1$
    is.  $\cos\theta_A=0$ is decoupled from the Standard Model, while
    $\cos\theta_A=1$ indicates that the $a_1$ is identical to the MSSM
    $A$.  In the bottom panels, dark (blue) = $m_{a_i}<2m_\tau$; medium grey
    (red) = $2m_\tau<m_{a_i}<8.4$ GeV; light grey (cyan) =
    $8.4{\rm GeV}<m_{a_i}<m_\Upsilon$.
 The plots are for $\tan \beta=10$ and $M_{1,2,3}=100,200,300$ GeV at scale
 $M_Z$.
The bottom left plot comes from simply scanning in $A_\lambda,A_\kappa$ holding
$\mu_{eff}=150$ GeV fixed.  The bottom right plot shows results for the $F<15$
scenarios among the orange-cross, i.e.\ $m_{a_i}<2m_b(pole)$, points of
Fig.~1 of Ref.\cite{Dermisek:2006py}. }
\end{figure}

Several of these measurements have now been performed including
$\Upsilon(1S) \to \chi \chi$\cite{Rubin:2006gc,Tajima:2006nc}; $\eta \to
\chi \chi$ and $\eta^\prime \to \chi \chi$\cite{Ablikim:2006eg}; and now
$J/\Psi \to \chi \chi$\cite{:2007ek}.

\subsection{Radiative Decay}
Radiative decay refers to meson decays into something visible as well as
something invisible.  This can be flavor changing, such as $b \to s \chi
\chi$, in which case this is a next-to-leading-order effect requiring a
loop of $W^\pm$ bosons to induce flavor changing.\cite{Bird:2006jd}  The
authors of Ref.\cite{Bird:2006jd} found that this radiative decay can be
as much as 50 times larger than the similar process radiating neutrinos.
\footnote{We prefer to avoid introducing flavor-changing couplings of a
Dark Matter mediator, as this is can introduce large corrections to the
CKM matrix.}  Other modes include $\Upsilon \to \gamma \chi \chi$ and
$J/\Psi \to \gamma \chi \chi$\cite{nmssmdm}.

In Fig.\ref{brgammachichi} we show the branching ratio of $J/\Psi$ and
$\Upsilon$ into $\gamma a_1$ in the NMSSM.\cite{Dermisek:2006py}  The
relevance of these plots for Dark Matter are that the $a_1$ may decay
into the neutralino if the bino mass $M_1$ is decreased to make this mode
kinematically allowed (without affecting this branching ratio).  This
can be done such that it is compatible with all collider constraints
including $Z \to invisible$ as described in Ref.\cite{nmssmdm}.

These branching ratios have little to do with the model assumptions of
the NMSSM and can be parameterized only with $\theta_A$ and $\beta$:
\begin{eqnarray}
    BR(\Upsilon \to \gamma a_1) &\propto& \cos \theta_A \tan \beta\\
    BR(J/\Psi   \to \gamma a_1) &\propto& \cos \theta_A \cot \beta,
\end{eqnarray}
so that experimentally, the only thing one needs worry about is that
$M_{a_1}$ is small enough that the mode is kinematically allowed, and
any limit can be interpreted in the $\cos \theta_A \tan \beta$ vs.
$M_{a_1}$ plane.  This covers a vast array of model space including any
model with a light Higgs having some singlet admixture.  Such Higgses
appear in Refs.\cite{Han:2004yd,nmssmdm,Barger:2006dh}

\section{Higgses}

Likewise, the existence of light Higgses can completely destroy the
observability of Standard Model Higgs signals at the LHC via decays such
as $h_2 \to h_1 h_1$ if it is dominant, where $h_2$ is a SM-like Higgs
and $h_1$ is a lighter, mostly-singlet Higgs.  Even if substantial
backgrounds at the LHC can be overcome, the LHC will be unable to get a
precise measurement of the lighter $h_1$ mass.  If $h_1$ decays to
$\tau^+ \tau^-$ the missing energy makes the mass measurement imprecise.
If the $h_1$ decays to charm, strange, or gluons this renders the
dominant Higgs decay entirely hadronic, and likely unobservable at the
LHC due to hadronic backgrounds.

By contrast, bottom and
charm factories can obtain precise measurements of the mass via the
energy of a recoiling photon in the process $\Upsilon \to \gamma
h_1$.\cite{Dermisek:2006py}  The $h_1$ may have branching fractions to
both Standard Model matter and Dark Matter.

The mode $\Upsilon \to \gamma H (A)$ was first suggested by
Wilczek.\cite{Wilczek:1977zn}  This mode is subject to significant
radiative and threshold corrections, a comprehensive list of which can
be found in Ref.\cite{hhg}.  It was vigorously pursued until about 1995,
when it became clear that the LEP accelerator, searching for SM or MSSM
Higgses in the Higgsstrahlung modes $e^+ e^- \to Z h$ and $e^+ e^- \to A
h$ was superior.  The best measurements on the Upsilon were made by
CLEO,\cite{cleolimits} however these remain about a factor of 10 away
from being sensitive to a Standard Model Higgs.  Existing data can reach
the sensitivity required, as CLEO and Belle have approximately 20 times
more data collected than that used in these limits.  It will be
necessary to reach and exceed the Standard Model limits to have
sensitivity to Higgses with some singlet admixture.

The LEP measurements told us that no new particles with masses below $M_Z$ have a
significant coupling to the $Z$.  However, they tell us little about
particles which have small coupling to the $Z$, and cannot rule out the
existence of light particles.  Particles with small $Z$ coupling are
still allowed and can have interesting couplings to Higgses and fermions.

It is perhaps surprising that a light Higgs could still exist at low
energies, and be compatible with all existing direct and indirect
limits.  However numerous studies have borne this out in a variety of
models.  All relevant experimental limits have been checked and light Higgses
remain consistent with them.  Some examples are: In the context of the Two
Higgs Doublet Model, $(g-2)_\mu$ (the anomalous magnetic moment of the muon)
was examined\cite{Krawczyk:2001pe,Larios:2002ha}, as well as BR$(b \to s
\gamma)$, $R_b$, $A_b$, BR$(\Upsilon \to A \gamma)$, BR$(\eta \to A
\gamma)$\cite{Larios:2002ha}. In the context of the NMSSM, BR$(\Upsilon \to
\gamma + X)$\cite{nmssmdm} was examined and found to be compatible.

We should note also that there is experimental evidence that this decay
exists, from considerations of the excess seen at LEP near $M_h=100$
GeV, fine tuning in the NMSSM,\cite{Dermisek:2005ar}, as well as some
anomalous events at the HyperCP experiment which seem to indicate a
$\sim 250$ MeV pseudo-scalar decaying to muons\cite{Mangano:2007gi} that
can be verified using radiative decays.

To allow a Higgs to be light, one must reduce its coupling to the $Z$
boson.  In the MSSM this is proportional to $\sin(\beta-\alpha)$ for the
CP-even state, and zero (at tree level) for the CP-odd state.  Thus, by
tuning the Higgs mixing angle $\alpha$ to be close to the ratio of the
vacuum expectation values $\tan \beta$, this can be achieved.  In the MSSM,
however, the relationships among masses, $\alpha$, and $\beta$ is too
constrained to allow only one of the Higgses to be lighter than $M_Z$
while simultaneously satisfying the Higgsstrahlung constraints.
Basically, one of the CP-even Higgses has a mass related to the CP-odd
Higgs, and the other is related to $M_Z$.  So one cannot bring the $h$
light while simultaneously keeping the $A$ heavy.  The $A$ becomes light
as well, and generates a large cross section for $e^+ e^- \to h A$.

This difficulty comes from the fact that there is not enough freedom in
the Higgs mass matrices, and as such is a theoretical constraint caused
by one's assumptions, and not experimental proof that there is no light
Higgs.  In models with more Higgs particles or more freedom in the Higgs
self-couplings, the $Z Z h$ coupling is more complex, and can be made
small.  The expansion of the Higgs sector in this manner is well
motivated from the need to break any extra gauge symmetries such as a
$U(1)^\prime$\cite{Han:2004yd} or $SU(2)_R$, or to solve the MSSM's
$\mu$ problem.\cite{nmssm}  Such particles may also generically be
associated with SUSY breaking.

In a more general Two Higgs Doublet Model (2HDM), small coupling to the
$Z$ can be achieved with light Higgses because $\beta$ and $\alpha$ are
essentially free parameters.  There remains some interesting parameter
space in the 2HDM accessible at $b$- and $c$- factories, but it is
small.\cite{Krawczyk:2001pe} 

Finally there now exists ``Gauge-Phobic Higgs'' models in which
electroweak symmetry breaking occurs by a combination of an elementary
Higgs and breaking by by boundary conditions in an extra
dimension.\cite{Cacciapaglia:2006mz}  In such models, the Higgses become
decoupled from the $Z$. Their mass again becomes a free parameter and
can be light, depending on how much of the symmetry breaking occurs due
to the Higgs and how much due to the extra dimension.

\section{The Future}

As the $b$-Factories come to the end of their lives, much attention has
been given to possible runs off the $\Upsilon(4S)$ resonance.  This is
an extremely promising idea.  A small amount (e.g. weeks to months) of
run time at a different energy may provide powerful physics results.
Spending that same time on the $\Upsilon(4S)$ will provide only a
negligible improvement over the already precise flavor physics results
returned by these machines.

The promising options for future runs are on the $\Upsilon(3S)$,
$\Upsilon(1S)$, $\Psi(2S)$, and $J/\Psi$.  We have argued for the
$\Upsilon(3S)$ due to the existence of the radiative decay $\Upsilon(3S)
\to \pi \pi \Upsilon(1S)$,\footnote{The $\Upsilon(2S)$ has this decay
mode also, as used in the invisible $\Upsilon$ search by
CLEO\cite{Rubin:2006gc}, however the pions are softer and more difficult
to reconstruct.} which can be used as a powerful constraint to
remove backgrounds for invisible searches\cite{McElrath:2005bp} and
Higgs searches\cite{Dermisek:2006py}.  In addition to the CLEO data
collected in the 1990's, Belle has already collected 2.9
fb${}^{-1}$\cite{Tajima:2006nc} on the $\Upsilon(3S)$.

No new studies have yet been published on the radiative decays $\Upsilon
\to \gamma + X$ or $J/\Psi \to \gamma + X$.  To improve on the
capabilities of the CLEO datasets when searching for rare decays, it is
necessary to further reject backgrounds.  The single photon signal
itself (ignoring the rest of the event) has sizeable backgrounds from
direct production of 3-body final states $f \bar{f} \gamma$ which is a
background to $\Upsilon \to \gamma a_1 \to \gamma f \bar{f}$.  This
argument was presented for $\Upsilon \to \gamma \tau^+ \tau^-$ in
Ref.\cite{Dermisek:2006py}, but holds just as well for other fermions.
Due to the high luminosity, BaBar and Belle have higher photon
backgrounds in general than CLEO did.

Therefore, for all searches described here, we believe that new runs on
the $\Upsilon(3S)$ and $\Psi(2S)$ will be the most significant.  A
Super-B factory can be even more powerful.\cite{Fullana:2007uq}

\bigskip 


\end{document}